\documentclass[aps,prb,twocolumn,floats,showpacs]{revtex4}

\usepackage{dcolumn}
\usepackage{graphicx}
\usepackage{epsfig}

\newcommand {\ket}[1]{|{#1}\rangle}

\def\mmat#1{\hbox{$\sf{#1}$}}

\begin{document}

% =============================================================================
\title{Non-Markovian dynamics in plasmon-induced spontaneous emission interference}

\author{I. Thanopulos$^{1}$}

\email{ithano@eie.gr}

\author{V. Yannopapas$^2$}

\email{vyannop@mail.ntua.gr}

\author{E. Paspalakis$^{3}$}

\email{paspalak@upatras.gr}

\affiliation{$^1$ Department of Optics and Optometry, T.E.I. of Western Greece, Aigio 251 00, Greece}
%\affiliation{$^1$ Department of Optics and Optometry, T.E.I. of Western Greece, Aigio 25100, Greece, and  Theoretical Physical Chemistry Institute, National Hellenic Research Foundation, Athens 11635, Greece}

\affiliation{$^2$ Department of Physics, National Technical University of Athens, Athens 157 80, Greece}

\affiliation{$^3$ Department of Materials Science, School of Natural Sciences,
University of Patras, Patras 265 04, Greece}

\begin{abstract}

We investigate theoretically the non-Markovian dynamics of a degenerate {\sf V}-type quantum emitter in the vicinity of a metallic nanosphere, a system that exhibits quantum interference in spontaneous emission due to the anisotropic Purcell effect. We calculate numerically the electromagnetic Green's tensor and employ the effective modes differential equation method for calculating the quantum dynamics of the emitter population, with respect
to the resonance frequency and the initial state of the emitter, as well as its distance from the nanosphere.
We find that the emitter population evolution varies between a gradually total decay and a partial decay combined with oscillatory population dynamics, depending strongly on the specific values of the above three parameters. Under
strong coupling conditions, coherent population trapping can be observed in this system. We compare our exact results with results when the flat continuum approximation for the modified by the metallic nanosphere vacuum is applied. We conclude that the flat continuum approximation is an excellent approximation
only when the spectral density of the system under study is characterized by non-overlapping plasmonic resonances.

\end{abstract}


\maketitle

% =============================================================================

\section{Introduction}

Metallic nanoparticles (MNPs) can confine light inside deep subwavelength
volumes in electromagnetic (EM) modes, which are called localized surface plasmons
(LSPs). This strong field localization leads to an enhancement of the light-matter
interaction in the vicinity of the MNP, featuring some very interesting physics in
a broad range of contexts such as optical antennas \cite{novotny11,maier11}, surface-enhanced Raman scattering \cite{schlucker14}, photovoltaics \cite{atwater10},
energy transfer in light-harvesting \cite{hulst16} and biosensing \cite{vollmer12}.

The presence of LSPs in metal-dielectric interfaces strongly modifies the density
of the EM modes in the surroundings, leading to strong modification of the lifetime
of quantum emitters (QE) placed close to the MNP. Moreover, when the distance
between the QE and the MNP surface is very small, theory predicts that coherent energy exchange between the
QE and the modified EM modes accompanied with non-Markovian dynamics may take place  \cite{tejedor10,hummer13,tejedor14,delga14,hakami14,hakami16}.
The above phenomena have been demonstrated mostly experimentally in the perturbative
(weak-coupling) regime wherein quantum dynamics is irreversible \cite{gittins02,sandog06,novotny06,Moerner09,andersen11,Hoang2015,Hoang2016}.
Nevertheless, recently strong coupling at room temperature between an organic molecule and
a plasmonic nanocavity has been demonstrated experimentally \cite{baumberg}, followed by
a detailed theoretical analysis \cite{garcia16}.

In this work we study theoretically the population dynamics of a {\sf V}-type QE coupled
electromagnetically to a MNP. The spontaneous emission of such a QE may exhibit interference
effects due the fact that the MNP affects differently the EM modes along mutually perpendicular
directions, known as the anisotropic Purcell effect \cite{Agarwal00a,Eversrev}.
This phenomenon has been studied in various photonic structures, including periodic dielectrics \cite{Li01a},
negative refractive index metamaterials \cite{Yang08a,Li09a,Xu10a}, metasurfaces \cite{Jha15a}, hyperbolic metamaterials \cite{Sun16}, as well as plasmonic nanostructures \cite{vasilis09,paspa11,Gu12a}.

\vspace*{1.cm}
\begin{figure}[h]
\centerline{\hbox{\epsfxsize=70mm\epsffile{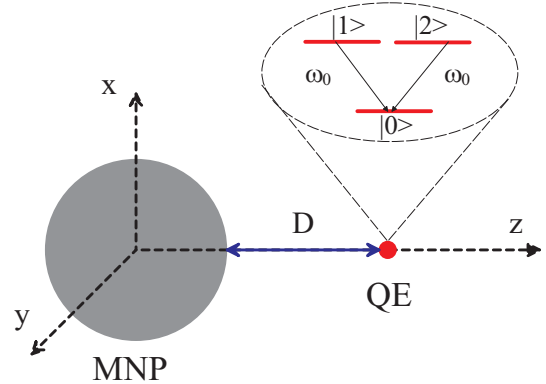}}}
\begin{center}
\caption{\label{fig1} (color online)  The configuration of the degenerate {\sf V}-type QE with
resonance frequency $\omega_0$, placed at a distance $D$ from the surface of a spherical MNP.}
\end{center}
\end{figure}

Here, in contrast to previous work \cite{Agarwal00a,Eversrev,Li01a,Yang08a,Li09a,Xu10a,Jha15a,Sun16,vasilis09,paspa11,Gu12a},
the interaction of the QE with the MNP is not restricted to the weak-coupling regime. We use a
quantum-mechanical formalism for the EM field excitations \cite{tejedor14,delga14,hakami14}, which goes
beyond the Markov approximation and incorporates the exact population dynamics by employing the effective
mode differential equation (EMDE) method \cite{emde}. The anisotropic enhancement of the spontaneous decay rates of a QE in the proximity of a MNP is calculated using an EM Green's tensor technique \cite{Yannopapas2007}. The purpose of the present work is to study in detail the crossover regime in which the QE population dynamics changes from irreversible to reversible for a system shown in Fig. \ref{fig1}, manifested by the emergence of non-decaying oscillations in the evolution of the QE population. Such oscillations signify the coherent exchange of energy between the QE and modified reservoir of EM modes due to the nearby MNP.

This paper is organized as follows: In Section \ref{theory} we present our methodology for calculating the spontaneous emission dynamics of a QE near a MNP. We then present and discuss the results for a {\sf V}-type system in Section \ref{vtype}. We also present relevant results for a two-level system in Section \ref{twolev}. Finally, we conclude our findings in Section \ref{concl}.

\section{Calculation of the spontaneous emission dynamics of a $V$-type QE near a MNP}
\label{theory}

We consider a degenerate {\sf V}-type QE at distance $D$ from the surface of a spherical MNP, as shown in 
Fig. \ref{fig1}. The origin of the coordinate system coincides with the center of the sphere and the QE lies on 
the z axis of the coordinate system. The Hamiltonian of the system \cite{tejedor14,hakami14}, is given by  
(we use $\hbar=1$ throughout this work)
\begin{eqnarray}
H &=& \sum_{i=1,2}\omega_0\hat{\bf \sigma}_{ii}+\int d\vec{r}\int_0^\infty d\omega~\omega~
\hat{\bf f}^{~\dagger}(\vec{r},\omega)\cdot \hat{\bf f}(\vec{r},\omega) \nonumber \\
&-&\sum_{i=1,2}\left[\hat{\bf \sigma}_{i0}\int_0^\infty d\omega \vec{\bf \mu}_{i0}\cdot
\hat{\bf E}(\vec{r},\omega)+ H.c.\right] \label{hamil}\,.
\end{eqnarray}
In Eq. (\ref{hamil}), $\hat{\bf f}(\vec{r},\omega)$, $\hat{\bf f}^{~\dagger}(\vec{r},\omega)$
stand for the bosonic vector field operators for the elementary excitations of the system, $\hat{\bf \sigma}_{ij}$ denotes the Pauli operator, $\vec{\bf \mu}_{10}= \mu\hat{\varepsilon}_{-}$ and $\vec{\bf \mu}_{20} = \mu\hat{\varepsilon}_{+}$ are
the dipole moments of the QE, where $\hat{\varepsilon}_\pm\equiv(\hat{\varepsilon}_z\pm i\hat{\varepsilon}_{x(y)})/\sqrt{2}$ describe the right-rotating ($\hat{\varepsilon}_{+}$)
and left-rotating ($\hat{\varepsilon}_{-}$) unit vectors ($\mu$ is taken to be real). Also,
$\omega_0$ stands for the resonance frequency between the two degenerate upper levels and the lower level of
the {\sf V}-type QE, with the energy of the lower level taken as zero.

The electric field vector operator $\hat{\bf E}(\vec{r},\omega)$ is given by
\begin{equation}\label{efield}
\hat{\bf E}(\vec{r},\omega)=\frac{i\omega^2}{\sqrt{\pi\epsilon_0}c^2}\int d\vec{s}
\sqrt{\mbox{Im}[\epsilon(\vec{s},\omega)]} \hat{\bf{G}}(\vec{r},\vec{s},\omega)\cdot
\hat{\bf f}(\vec{s},\omega) \, ,
\end{equation}
with $\hat{\bf{G}}(\vec{r},\vec{s},\omega) $ being the dyadic EM Green's tensor defined as
\begin{equation}\label{helmhol}
\nabla\times\nabla\times\hat{\bf{G}}(\vec{r},\vec{s},\omega) -
\frac{\epsilon(\vec{r},\omega)\omega^2}{c^2}\hat{\bf{G}}(\vec{r},\vec{s},\omega)
= \hat{\bf{I}}\delta(\vec{r}-\vec{s}~)\,.
\end{equation}
In Eqs. (\ref{efield}) and  (\ref{helmhol}), $\hat{\bf{I}}$ is the unit dyad (unit tensor),
$\epsilon(\vec{r},\omega)$ is the spatially and frequency dependent complex dielectric function of the MNP, and
$c$ is the speed of light in the vacuum.

The state of the system is given by
\begin{eqnarray}
\ket{\Psi(t)}&=&c_1(t) e^{-i\omega_0t}\ket{1;0_\omega} + c_2(t)
e^{-i\omega_0t}\ket{2;0_\omega} + \nonumber \\
&&\int d\vec{r}\int d\omega~ C(\vec{r},\omega,t) e^{-i\omega t}
\ket{0;1_{\vec{r},\omega}} \, ,
\end{eqnarray}
where $\ket{0;1_{\vec{r},\omega}}\equiv \hat{f}_\lambda^{~\dagger}(\vec{r},\omega)\ket{0;0_\omega}$. Here, 
$\ket{n;a} = |n\rangle \otimes |a\rangle$, where $|n\rangle$ $(n=0, 1, 2)$ denotes the quantum states of the 
V-type system (see Fig. 1) and $|a\rangle$ denotes the photonic states (states of the modified EM vacuum), with 
$\ket{0_\omega}$ meaning zero and $\ket{1_{\vec{r},\omega}}$ one photon states.
The application of the time-dependent Schr\"odinger equation $i\ket{\dot{\Psi}(t)}=H \ket{\Psi(t)}$ yields
a set of differential equations that can be formally integrated, resulting in two integro-differential
equations for $c_1(t)$ and $c_2(t)$, given by 
\begin{equation}\label{eq0a}
\dot{c}_1(t) = i\int_0^t dt^\prime
\Big(K^{11}(t-t^\prime) c_1(t^\prime) + K^{12}(t-t^\prime) c_2(t^\prime) \Big) \, , \\
\end{equation}
\begin{equation}\label{eq0b}
\dot{c}_2(t) = i\int_0^t dt^\prime
\Big( K^{21}(t-t^\prime) c_1(t^\prime) + K^{22}(t-t^\prime) c_2(t^\prime) \Big) \, , \\
\end{equation}
where 
\begin{eqnarray}
K^{ij}(\tau)  & = &  i \frac{e^{i\omega_0\tau}}{\pi}
\int_0^\infty {\cal G}^{ij}(\omega) e^{-i\omega\tau} d\omega \label{kernel} \\ 
{\cal G}^{ij}(\omega) & \equiv & 
\frac{\omega^2}{\epsilon_0 c^2}\Big[\vec{\mu}^{\dagger}_{i0} \cdot \mbox{Im}
[\hat{\bf G}(\vec{r}_0,\vec{r}_0,\omega)] \cdot \vec{\mu}_{j0}\Big] \, ,
\end{eqnarray}
with $\tau\equiv t-t^\prime$ and $i,j=1,2$.

In general, the MNP affects differently the EM modes along the tangential ($\parallel$) and the
radial ($\perp$) directions, corresponding to a QE with a transition dipole oriented parallel and
perpendicular to the surface of the MNP, respectively. In our case, the radial direction coincides with the $z$ axis, and the tangential direction is parallel to the $x$ and $y$ axes,
according to Fig. 1. In such a case, we have $(i,j=1,2)$ 
\begin{equation}
K^{ii}(\tau)=K^+(\tau) =  ie^{i\omega_0\tau}\int_0^\infty J^+(\omega) e^{-i\omega\tau} d\omega \, ,
\end{equation}
\begin{equation}
K^{ij}(\tau)=K^-(\tau) = ie^{i\omega_0\tau} \int_0^\infty J^-(\omega) e^{-i\omega\tau} d\omega  \, ,i\ne j
\end{equation}
with
\begin{eqnarray}\label{dens}
J^\pm(\omega) & = &\frac{1}{2\pi}\frac{\omega^2\mu^2}{\epsilon_0 c^2} \mbox{Im}
[\hat{\bf G}_\perp(\vec{r}_0,\vec{r}_0,\omega)\pm\hat{\bf G}_\parallel(\vec{r}_0,\vec{r}_0,\omega)] 
\nonumber\\
&\equiv&J^\perp(\omega)\pm J^\parallel(\omega) \, ,
\end{eqnarray}
where $\hat{\bf G}_\perp\equiv\hat{\bf G}_{zz}$ and $\hat{\bf G}_\parallel\equiv\hat{\bf G}_{xx(yy)}$.
The decay rate $\Gamma(\omega)$ in the presence of a MNP is given by \cite{welsch02}
\begin{equation}
\Gamma^{k}(\omega)=\frac{2\mu^{2}\omega^2}{\epsilon_0 c^2}\mbox{Im}[\hat{\bf G}_{k}(\vec{r}_0,\vec{r}_0,\omega)] \, , \quad k=\perp,\parallel \, ,
\end{equation}
and we thus obtain
\begin{equation} \label{jk}
J^k(\omega)=\frac{1}{2\pi}\frac{\Gamma^k(\omega)}{2}=\frac{1}{2\pi}\frac{\lambda^k(\omega,D) \Gamma_0(\omega)}{2} \, ,
\end{equation}
with $\Gamma_0(\omega)$ being the decay rate of the QE in free-space, and $\lambda^k(\omega,D)$ is the directional ($k=\perp,\parallel$) enhancement factor of the free-space decay rate due to the presence of the MNP at distance $D$ from the QE.  Due to the anisotropy, the enhancement factors $\lambda^\parallel(\omega,D)$ 
and $\lambda^\perp(\omega,D)$  are not equal, in general. We further write
\begin{equation}
\Gamma_0(\omega)= \frac{\omega_0^3\mu^2}{3\pi\epsilon_0 c^3}  (\frac{\omega}{\omega_0})^3 \equiv   \Gamma_0(\omega_0)(\frac{\omega}{\omega_0})^3 \, ,
\end{equation}
since in free-space $\mbox{Im}[\hat{\bf G}_0(\vec{r}_0,\vec{r}_0,\omega)] =\frac{\omega}{6\pi c} \hat{\bf{I}}$ 
is valid \cite{hakami16,welsch02}, with $\hat{\bf G}_0(\vec{r}_0,\vec{r}_0,\omega)$ being the 
free-space Green's tensor.
We now obtain
\begin{eqnarray}\label{with}
K^\pm(\tau) & = & i \int_0^\infty J^\pm(\omega) e^{-i(\omega-\omega_0)\tau}d\omega \, , \\
J^\pm (\omega)&\equiv& \frac{\Gamma_0(\omega_0)}{2\pi}    \frac{\lambda^\perp(\omega,D)\pm
\lambda^\parallel(\omega,D)}{2} (\frac{\omega}{\omega_0})^3 \nonumber \\
&\equiv& J^{rad}(\omega)\pm J^{tan}(\omega) \, ,
\end{eqnarray}
with $\Gamma_0(\omega_0)=1/\tau_0$, and $\tau_0$ standing for the free-space decay time of the QE. Taking into account that $\exp(-i(\omega-\omega_0)\tau)$ contributes mainly around $\omega_0$, we can write 
$(\frac{\omega}{\omega_0})^3\approx 1$, introducing thus the flat continuum approximation (FCA), and  
obtaining 
\begin{eqnarray}\label{without}
K^\pm(\tau) & \approx & K^\pm_{FCA}(\tau) \\
K^\pm_{FCA} & = &  i \int_0^\infty J^\pm_{FCA}(\omega)
e^{-i(\omega-\omega_0)\tau}d\omega \, , \\
J^\pm_{FCA}(\omega)&\equiv&
\frac{\Gamma_0(\omega_0)}{2\pi}   \frac{\lambda^\perp(\omega,D)\pm
\lambda^\parallel(\omega,D)}{2} \nonumber \\
&\equiv&  J^{rad}_{FCA}(\omega)\pm J^{tan}_{FCA}(\omega) \, .
\end{eqnarray}

We can now expand (to any given accuracy) $K^\pm(\tau)$  as a sum of $M$
exponential terms at {\it effective mode} frequencies $\omega_i$ ($i=1,\ldots,M)$ \cite{emde},
\begin{equation}\label{expansio}
K^\pm(\tau) = i e^{i\omega_0\tau} \sum_{i=1}^{M} W_i^\pm e^{-i\omega_i\tau} \, .
\end{equation}
Then, the equations for the two upper states of the QE are given by
\begin{equation}\label{eq1}
\dot{c}_\nu(t) = i\sum_{i=1}^M  e^{i(\omega_0-\omega_i)t} {\cal J}_i^\nu(t)\,,~ \nu=1,2 \, 
\end{equation}
where we have introduced  new variables given by
\begin{equation}\label{eq2a}
{\cal J}_i^1(t)  =  i\int_0^t \Big(W_i^+c_1(t^\prime) + W_i^-c_2(t^\prime)\Big) e^{-i(\omega_0-\omega_i)t^\prime}dt^\prime \, ,
\end{equation}
\begin{equation}\label{eq2b}
{\cal J}_i^2(t)  =
i\int_0^t \Big(W_i^-c_1(t^\prime) + W_i^+c_2(t^\prime)\Big) e^{-i(\omega_0-\omega_i)t^\prime}dt^\prime \, ,
\end{equation}
for which holds
\begin{eqnarray}
\dot{\cal J}^1_i(t) &=& i (W_i^+c_1(t) + W^-_ic_2(t))  e^{-i(\omega_0-\omega_i)t} \label{eq3a} \, , \\
\dot{\cal J}^2_i(t) &=& i (W_i^-c_1(t) + W^+_ic_2(t))e^{-i(\omega_0-\omega_i)t} \label{eq3b} \, .
\end{eqnarray}
In this way, we have transformed the two integro-differential equations, Eqs. (\ref{eq0a}), (\ref{eq0b}), into a set
of $2(M+1)$ effective modes (ordinary) differential equations (EMDE) with constant coefficients, given by
Eqs. (\ref{eq1}), (\ref{eq3a}) and (\ref{eq3b}). Denoting $\tilde{C}_\nu(t)\equiv\exp(-i\omega_0t)c_\nu(t)$ and
$\tilde{{\cal J}}_i^\nu(t)\equiv\exp(-i\omega_i t) {\cal J}_i^\nu(t)$, with $\nu=1,2$ and $i=1,\ldots,M$, we 
obtain Eqs. (\ref{eq1}), (\ref{eq3a}) , and (\ref{eq3b}) in a matrix form
\begin{equation}
{\bf \dot{\tilde{C}}}(t)=
i\,{\mmat {\bf \tilde{H}}}\cdot{\bf \tilde{C}}(t) \, ,
\label{matreq2}
\end{equation}
where ${\bf \tilde{C}}(t)=(\tilde{C}_{1}(t),\tilde{C}_{2}(t), \tilde{{\cal
J}}_{1}^1(t),...,\tilde{{\cal J}}_{M}^1(t), \tilde{{\cal
J}}_{1}^2(t),...,\tilde{{\cal J}}_{M}^2(t))^{\mmat{T}}$ \linebreak
 $\equiv(\tilde{C}_{1}(t),\tilde{C}_2(t), \tilde{{\cal
J}}^{(1)}(t),...,\tilde{{\cal J}}^{(2M)}(t))^{\mmat{T}}$, with
all $\tilde{{\cal J}}^{(i)}(0)=0$.
Here the superscript ${\mmat{T}}$ is the transpose operation and
$\mmat{\bf \tilde{H}}$ is a time-independent complex matrix,  given by
\begin{equation}
\mmat {\bf\tilde{H} }  =  \left(
\begin{array}{ll}
\mmat{{\cal S}} &
\mmat{{\cal Y}} \\
\mmat{\cal Z} &
\mmat{ {\cal R}} \\
\end{array} \right),\,\,\,
\mmat {\cal Z} =  \left(
\begin{array}{l}
\mmat{W}^{(1)} \\
... \\
\mmat{W}^{(2M)} \\
\end{array}
\right),\,\,
\mmat {\cal Y} =  \left(
\begin{array}{c}
\mmat{\Omega}^{(1)} \\
\mmat{\Omega}^{(2)} \\
\end{array}
\right)\,.
\label{hef4}
\end{equation}\
In Eq. (\ref{hef4}),  we have ${\cal S}_{k,k^\prime}=\delta_{k,k^\prime}\omega_0$, $(k,k^\prime=1,2)$,
and ${\cal R}_{i,i^\prime}={\cal R}_{i+M,i^\prime+M}=\delta_{i,i^\prime}\omega_i$, $(i,i^\prime=1,\ldots,M)$,
for the diagonal blocks. Furthermore, the $2M$ rows of $\mmat{\cal Z}$ are defined as
$\mmat{W}^{(\eta)}=(W_\eta^+,W_\eta^-)$ and
$\mmat{W}^{(M+\eta)}=(W_\eta^-,W_\eta^+)$, for $\eta=1,\ldots,M$.
Lastly, the two rows of $\mmat{\cal Y}$ are defined as
$[\mmat{\Omega}^{(1)}]_{1,m}=[\mmat{\Omega}^{(2)}]_{2,n}=1$,  for
$3\le m\le 2+M $ and  $3+M\le n \le 2(M+1) $, with 
$[\mmat{\Omega}^{(1)}]_{1,m}=[\mmat{\Omega}^{(2)}]_{1,n}=0$, for $m$ and $n$ otherwise.
Given these definitions we write the solution of Eq. (\ref{matreq2}) as,
\begin{equation}\label{matreq3}
{\bf \tilde{C}}(t)= \sum_{j=1}^{M^\prime} {\bf \Lambda}_j e^{i\lambda_j t}\tilde{C}_j^0 \, ,
\end{equation}
where $M^\prime=2(M+1)$ is the number of $\lambda_j$ eigenvalues
and ${\bf \Lambda}_j$ eigenvectors of ${\bf\tilde{H}}$. 
The vector of complex coefficients $\tilde{C}^0_j$ is given as
${\bf \tilde{C}^0}={\bf L}^{-1}\cdot{\bf \tilde{C}}(0)$.
The columns of the ${\bf L}\equiv[{\bf \Lambda}_1,..., {\bf \Lambda}_{M^\prime}]$
matrix are the eigenvectors of ${\bf \tilde{H}}$.
We note that since ${\bf\tilde{H}}$
is not Hermitian, ${\bf L}^{-1}\ne{\bf L}^\dagger$.

\begin{figure}[h]
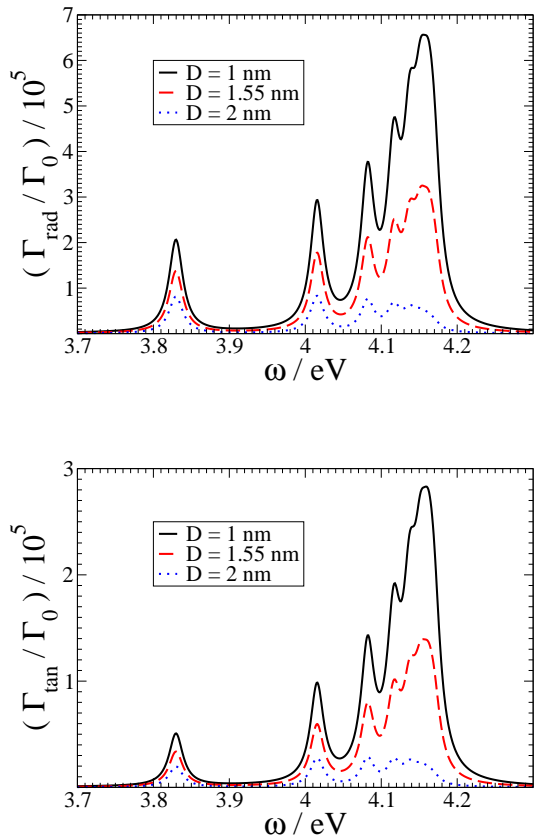

\centerline{\hbox{\epsfxsize=70mm\epsffile{fig2a.eps}}}
\vspace*{1.cm}
\centerline{\hbox{\epsfxsize=70mm\epsffile{fig2b.eps}}}
\begin{center}
\caption{\label{fig2} (color online)  Radial $\lambda^\perp(\omega,D)\equiv\Gamma_{rad}/\Gamma_0$
(top panel) and tangential  $\lambda^\parallel(\omega,D)\equiv \Gamma_{tan}/\Gamma_0$ (bottom panel)
enhancement factors as function of the distance $D$ of the QE from the
MNP and EM mode frequency $\omega$. }
\end{center}
\end{figure}

The EM response of the 5 nm silver sphere used in this study is described by a Drude dielectric function,
$\epsilon_m(\omega)=\epsilon_{m,\infty}-\omega_p^2/(\omega^2+i\omega\gamma)$, characterized by its
plasma frequency $\omega_p=9.176$ eV, high-frequency component $\epsilon_{m,\infty}=3.718$ eV, and
Ohmic losses $\gamma=0.021$ eV. The radial and tangential enhancement factors of the free-space decay 
rate of the QE due to the presence of the MNP as a function of frequency $\omega$ at distance $D$, 
calculated using an EM Green's tensor technique \cite{Yannopapas2007}, are presented in Fig. \ref{fig2}.
\vspace*{1.cm}

We note that our results below are converged with respect to the number of effective modes $M$ included in
the dynamics;  we use $M=1001$ in the energy range 3.5-4.5 eV. We also note that our results
on {\sf V}-type and two-level systems presented below are obtained using the $K^\pm(\tau)$ kernel as defined
in Eq. (\ref{with}) unless otherwise stated.

\section{Dynamics of a {\sf V}-type QE near a MNP}
\label{vtype}

In this work, we investigate the $|c_1(t)|^2$ and  $|c_2(t)|^2$ time evolution for  state-of-the-art QEs
with transition frequencies in the optical regime, such as quantum dots (QD) and J-aggregates (J-AGR), with
free-space decay time $\tau_0$ $\approx$ 4 ns \cite{akimov07} and $\approx$ 70 ps \cite{knoester90}, respectively; correspondingly, we often use the terms QD and J-AGR for referring to QEs with such free-space decay times below.  In each case, we study the dynamics for different initial states of the QE. Moreover, for each QE, we  consider two resonance frequencies, $\omega_0=3.84$ eV and $\omega_0=4.16$ eV, the frequencies corresponding to (relatively) small and large enhancement factors $\lambda^\parallel(\omega_0,D)$ and
$\lambda^\perp(\omega_0,D)$, respectively, shown in Fig. \ref{fig2}. We also define the effective decay time of a QE with resonance frequency $\omega_0$ and free-space decay rate $\tau_0$ at distance $D$ from 
the MNP surface as $\tau_{eff}\equiv(\tau_{eff}^{rad}+\tau_{eff}^{tan})/2$, given by the average of the corresponding radial $\tau_{eff}^{rad}\equiv \tau_0/\lambda_\perp(\omega_0,D)$ and tangential $\tau_{eff}^{tan}\equiv\tau_0/\lambda_\parallel(\omega_0,D)$
effective decay times.

\subsection{QD dynamics at $\omega_0=3.84$ eV}

\vspace*{1.cm}
\begin{figure}[h]
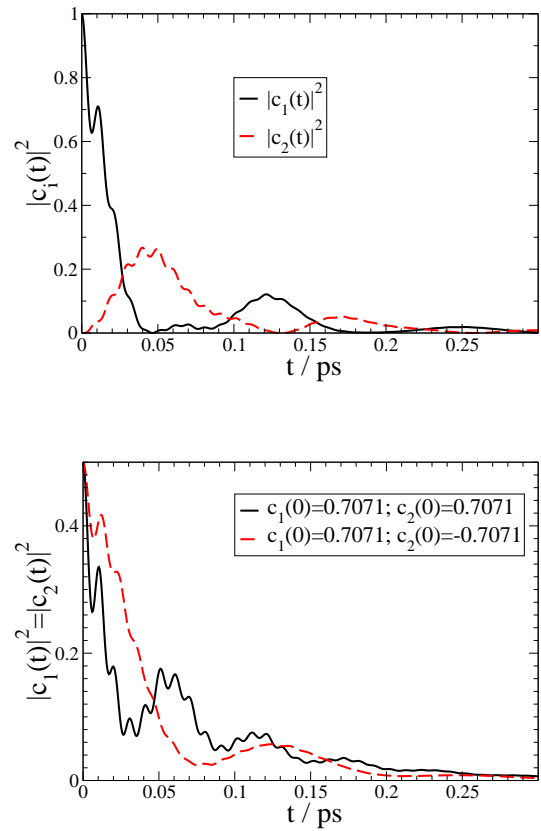

\centerline{\hbox{\epsfxsize=70mm\epsffile{fig3a.eps}}}
\vspace*{1.cm}
\centerline{\hbox{\epsfxsize=70mm\epsffile{fig3b.eps}}}
\begin{center}
\caption{\label{fig3} (color online) Top: Population evolution of $\ket{1}$ (black solid curve) and $\ket{2}$ (red  dashed curve) of a QD with $\omega_0=3.84$ eV at $D=1$ nm. Bottom: Population evolution of $\ket{1}$  of 
the QD with $\omega_0=3.84$ eV at $D=1$ nm for the initial states, 
$\ket{\Psi(0)}=\sqrt{0.5}\ket{1}+\sqrt{0.5}\ket{2}$ (black  solid curve) and 
$\ket{\Psi(0)}=\sqrt{0.5}\ket{1}-\sqrt{0.5}\ket{2}$ (red dashed curve).}
\end{center}
\end{figure}

The population dynamics of a QD, i.e. a QE with $\tau_0=4$ ns, with $\omega_0=3.84$ eV at $D=1$ nm for different initial states is shown in Fig. \ref{fig3}. In the top panel of this figure, the time-evolution of
$|c_1(t)|^2$ (black solid curve) and $|c_2(t)|^2$ (red dashed curve) is shown for the initial state $\ket{\Psi(0)}=\ket{1}$. We find that the initial population in state $\ket{1}$ decays rapidly into the modified EM modes, while there is a partial population transfer from state $\ket{1}$ into state $\ket{2}$ within about 50 fs. Partial revivals of the populations in $\ket{1}$ and $\ket{2}$ at 125 fs and 175 fs, respectively, are further observed, before all population finally ends up in the modified EM continuum after about 250 fs. The effective decay time for a QE with 
$\tau_0=4$ ns, as here, is $\tau_{eff}\approx$ 89 fs ($\tau_{eff}^{rad}\approx35$ fs and $\tau_{eff}^{tan}\approx143$ fs) considering that $\lambda_\perp(3.84,1)\approx 1.13\cdot10^5$ and 
$\lambda_\parallel(3.84,1)\approx 2.8\cdot10^4$ as shown in Fig.  \ref{fig2}. The time-evolution of state 
$\ket{1}$ shown in the top panel of Fig. \ref{fig3}, at early times, is practically a decay, characterised primarily 
by the decay time $\tau_{eff}^{rad}$, instead of $\tau_{eff}$, indicating that the clearly non-Markovian dynamics 
are dominated by the radial part (along the $z$ axis) of the modified by the MNP spectral density of the EM modes. Moreover, small oscillations on the population evolution dynamics of $\ket{1}$ and $\ket{2}$, which are discernible also in the top panel of Fig. \ref{fig3}, further point to a rather moderate non-Markovian character.
There is also a partial oscillatory exchange of population between state $\ket{1}$ and state $\ket{2}$ via the modified EM modes as indicated by a phase difference, which we define as $\pi$, since, when one is at a (local) maximum the other is at a (local) minimum, and vice versa. 

In the bottom panel of Fig. \ref{fig3} we present the population dynamics for $\ket{1}$ and $\ket{2}$ for the symmetric initial state $\ket{\Psi(0)}=\sqrt{0.5}\ket{1}+\sqrt{0.5}\ket{2}$ (black solid curve) and the antisymmetric initial state $\ket{\Psi(0)}=\sqrt{0.5}\ket{1}-\sqrt{0.5}\ket{2}$ (red dashed curve), denoted by SIS and AIS, respectively. We observe that the population evolution $|c_1(t)|^2$, which in case of SIS and AIS equals to $|c_2(t)|^2$, decays fast into the modified EM modes for both initial conditions; however, the explicit dynamics depends on the initial state. In case of a SIS, the population decay of $\ket{1}$ is slower than in case of an AIS; in addition, in the latter case, the non-Markovian character of the underlying dynamics is apparently stronger. The differences in the population evolution of state $\ket{1}$ and state $\ket{2}$ in case of a SIS and an AIS clearly indicate  that quantum interference effects affect the underlying dynamics, as well as the strength of the non-Markovian character on it. Furthermore, we observe that the population dynamics with a SIS and an AIS, at early times, can be understood as decays, with decay times $\tau_{eff}^{tan}\approx145$ fs and $\tau_{eff}^{rad}\approx35$ fs, respectively; apparently, the dynamics with a SIS is dominated by the tangential Purcell effect and the dynamics with an AIS is dominated by the radial Purcell effect. 

\subsection{J-AGR dynamics at $\omega_0=3.84$ eV}

\begin{figure}[h]
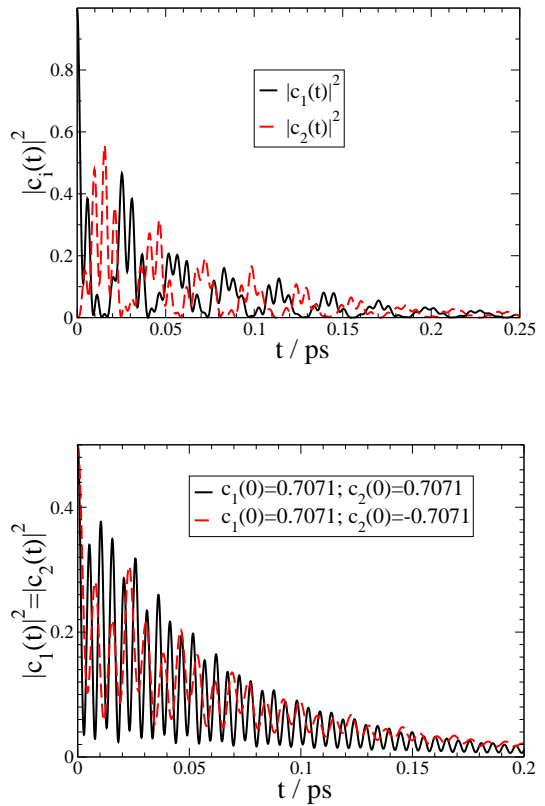

\centerline{\hbox{\epsfxsize=70mm\epsffile{fig4a.eps}}}
\vspace*{1.cm}
\centerline{\hbox{\epsfxsize=70mm\epsffile{fig4b.eps}}}
\begin{center}
\caption{\label{fig4} (color online) Top: Population evolution of $\ket{1}$ (black solid curve) and $\ket{2}$ (red  dashed curve) of a J-AGR with $\omega_0=3.84$ eV at $D=2$ nm. Bottom: Population evolution of $\ket{1}$  
of the J-AGR  with $\omega_0=3.84$ eV at $D=2$ nm for the initial states, $\ket{\Psi(0)}=\sqrt{0.5}\ket{1}+\sqrt{0.5}\ket{2}$ (black  solid curve) and $\ket{\Psi(0)}=\sqrt{0.5}\ket{1}-\sqrt{0.5}\ket{2}$
(red dashed curve).}
\end{center}
\end{figure}

We now focus on the dynamics of a J-AGR, i.e. a QE with $\tau_0=70$ ps, with resonance frequency
$\omega_0=3.84$ eV at $D=2$ nm, and corresponding $\tau_{eff}\approx 4$ fs ($\tau_{eff}^{rad}\approx1.6$ fs 
and $\tau_{eff}^{tan}\approx6.5$ fs), after taking into account that
$\lambda_\perp(3.84,2)\approx 4.40\cdot10^4$ and $\lambda_\parallel(3.84,2)\approx 1.07\cdot10^4$ as
shown in Fig.  \ref{fig2}. The population evolution of states $\ket{1}$ and state $\ket{2}$ of such a system is presented in Fig. \ref{fig4}.

\vspace*{1.cm}

In the top panel of Fig. \ref{fig4}, the evolution of $|c_1(t)|^2$ (black solid curve) and $|c_2(t)|^2$
(red dashed curve) for initial state $\ket{\Psi(0)}=\ket{1}$ are shown. We find that both $|c_1(t)|^2$ and 
$|c_2(t)|^2$ decay totally within 200-250 fs. In the meanwhile, though, there is partial oscillatory exchange of population between state $\ket{1}$ and state $\ket{2}$ via the modified EM modes as indicated by the phase difference of $\pi$. Moreover, the evolution of $|c_1(t)|^2$ and $|c_2(t)|^2$ point to a rather strong 
non-Markovian character of the underlying dynamics, as there are very distinct oscillations on the top of the 
overall gradual decay of the population of these states. 
The population evolution of $|c_1(t)|^2$ and $c_2(t)|^2$ for a SIS (black solid curve) and an AIS (red dashed curve) are shown in the bottom panel of Fig. \ref{fig4}. We observe that, in both cases, the population of state 
$\ket{1}$ and $\ket{2}$ decay totally in the modified EM mode continuum within about 200 fs. Furthermore, in both cases, the overall population decay occurs with strong, gradually decreasing in magnitude, oscillations of population transfer between states $\ket{1}$ and $\ket{2}$ and the modified EM continuum simultaneously. However, the density of oscillations within a given time interval is lower in case of a AIS than in case of a SIS, which points to destructive quantum interference effects in the underlying strong non-Markovian population dynamics in case of an AIS. We note that a closer look at the population evolution up to 5 fs in both panels of Fig. \ref{fig4} (not shown here) shows that the decay time in each case is close to the corresponding 
$\tau_{eff}^{rad}$, instead of $\tau_{eff}$, which indicates, that the radial part of spectral density is 
dominating the underlying non-Markovian population dynamics here, as in the case of a QD with the same 
$\omega_0$ shown Fig. \ref{fig3}.

\vspace*{1.cm}
\begin{figure}[h]
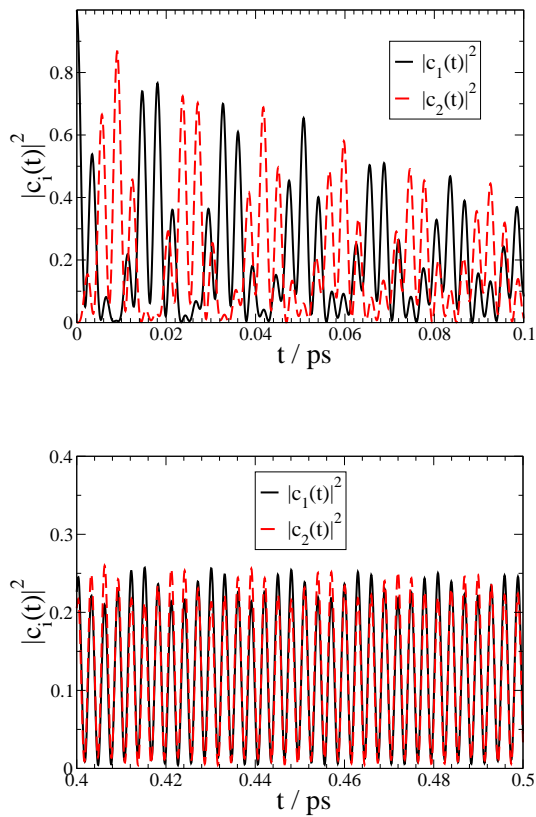

\centerline{\hbox{\epsfxsize=70mm\epsffile{fig5a.eps}}}
\vspace*{1.cm}
\centerline{\hbox{\epsfxsize=70mm\epsffile{fig5b.eps}}}
\begin{center}
\caption{\label{fig5} (color online) Top: Population evolution of $\ket{1}$ (black solid curve) and $\ket{2}$ (red  dashed curve) of a J-AGR with $\omega_0=3.84$ eV at $D=1.55$ nm up to 100 fs. Bottom: Population evolution of 
$\ket{1}$ (black solid curve) and $\ket{2}$ (red  dashed curve) of a J-AGR with $\omega_0=3.84$ eV at 
$D=1.55$ nm, at later times (400-500 fs).}
\end{center}
\end{figure}

In Fig. \ref{fig5}, we study the population evolution of states $\ket{1}$ and $\ket{2}$ for a J-AGR with
$\omega_0=3.84$ eV at $D=1.55$ nm, with $\tau_{eff}\approx2.3$ fs ($\tau_{eff}^{rad}\approx0.9$ fs 
and $\tau_{eff}^{tan}\approx3.8$ fs), taking into account that
$\lambda_\perp(3.84,1.55)\approx 7.50\cdot10^4$ and $\lambda_\parallel(3.84,1.55)\approx 1.85\cdot10^4$ 
as shown in Fig.  \ref{fig2}.  In the top panel of Fig. \ref{fig5}, the time evolution of $|c_1(t)|^2$ 
(black solid curve) and $|c_2(t)|^2$ (red dashed curve) with initial state $\ket{\Psi(0)}=\ket{1}$ are shown  
during the first 100 fs. At the very early stage of the population evolution  of state $\ket{1}$ (not shown here), 
the dynamics is practically a fast decay with $\tau_{eff}$; however, very quickly the strongly non-Markovian character of the underlying dynamics manifests itself as an oscillatory transfer of population between the 
states $\ket{1}$ and $\ket{2}$, with the modified EM mode continuum as an intermediate, as clearly indicated 
by the $\pi$-phase difference between $|c_1(t)|^2$ and $|c_2(t)^2$. The initial population of the {\sf V}-type 
system is decreased to about half within the first 100 fs.

\vspace*{1.cm}
\begin{figure}[h]
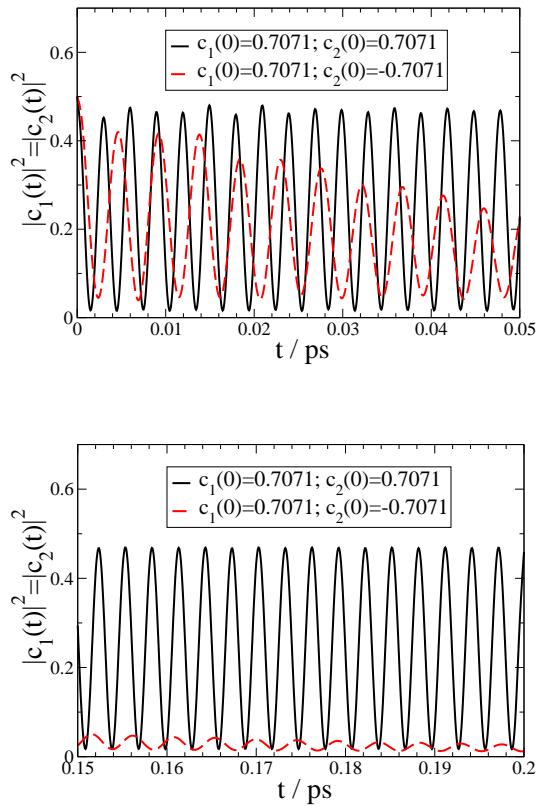

\centerline{\hbox{\epsfxsize=70mm\epsffile{fig6a.eps}}}
\vspace*{1.cm}
\centerline{\hbox{\epsfxsize=70mm\epsffile{fig6b.eps}}}
\begin{center}
\caption{\label{fig6} (color online) Top: Population evolution of $\ket{1}$  of a 
J-AGR  with $\omega_0=3.84$ eV at $D=1.55$ nm for the initial states, $\ket{\Psi(0)}=\sqrt{0.5}\ket{1}+\sqrt{0.5}\ket{2}$ (black  solid curve) and $\ket{\Psi(0)}=\sqrt{0.5}\ket{1}-\sqrt{0.5}\ket{2}$
(red dashed curve) up to 50 fs. Bottom: Population evolution of $\ket{1}$  of a 
J-AGR  with $\omega_0=3.84$ eV at $D=1.55$ nm for the initial states, $\ket{\Psi(0)}=\sqrt{0.5}\ket{1}+\sqrt{0.5}\ket{2}$ (black  solid curve) and $\ket{\Psi(0)}=\sqrt{0.5}\ket{1}-\sqrt{0.5}\ket{2}$
(red dashed curve) at later times (150-200 fs).}
\end{center}
\end{figure}

In the bottom panel of Fig. \ref{fig5} we present the population evolution $|c_1(t)|^2$ (black solid curve) and
$|c_2(t)|^2$ (red dashed curve) at later times, in the time interval 400 - 500 fs. We now find that the 
time-evolution of the population has reached a steady state, indicating a population exchange between the 
{\sf V}-type system and the modified EM mode continuum, which amounts in total to $\approx$ 50\% of the 
initial population. Interestingly, the $\pi$-phase shift between $|c_1(t)|^2$ and $|c_2(t)|^2$, which existed 
at earlier times of the dynamics, shown in the top panel of this figure, is now lost. The population exchange now 
has now become practically a transfer of population between the states of the {\sf V}-type system and the 
modified EM continuum, which occurs simultaneously for both states.

In Fig. \ref{fig6}, we investigate the influence of the initial state on the population evolution of states $\ket{1}$ and $\ket{2}$ for a J-AGR with $\omega_0=3.84$ eV at $D=1.55$ nm. In the top panel of this figure, we present
the time evolution of $|c_1(t)|^2$ and $|c_2(t)|^2$ for a SIS (black solid curve) and an AIS (red dashed curve) for the first 50 fs. We find that the population evolution of states $\ket{1}$ and $\ket{2}$ are strongly 
dependent on the initial state, which indicates significant quantum interference effects in the underlying 
non-Markovian dynamics.
In case of a SIS, the $|c_1(t)|^2$ ($=|c_2(t)|^2$) evolution shows a steady oscillatory exchange of population between the {\sf V}-type system and the modified EM mode continuum. In case of an AIS, the situation is strikingly different; the $|c_1(t)|^2$ evolution indicates an oscillatory gradually decreasing in magnitude transfer of the initial population of the {\sf V}-type system into the modified EM mode continuum. At later times, as shown in the bottom panel of Fig. \ref{fig6}, a marginal, only about 5\% of the initial population in total, oscillatory population transfer between states $\ket{1}$ and $\ket{2}$ and the modified EM modes, is still observable.

\subsection{QD dynamics at $\omega_0=4.16$ eV}

\vspace*{1.cm}
\begin{figure}[h]
\centerline{\hbox{\epsfxsize=70mm\epsffile{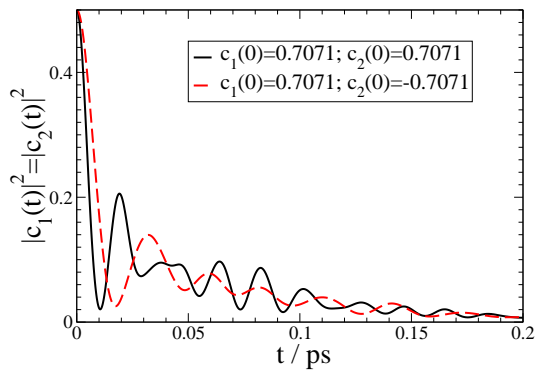}}}
\begin{center}
\caption{\label{fig7} (color online) Population evolution of $\ket{1}$  and $\ket{2}$ of a QD with
$\omega_0=4.16$ eV at $D=1$ nm for the initial states, $\ket{\Psi(0)}=\sqrt{0.5}\ket{1}+\sqrt{0.5}\ket{2}$
(black  solid curve) and $\ket{\Psi(0)}=\sqrt{0.5}\ket{1}-\sqrt{0.5}\ket{2}$ (red dashed curve).}
\end{center}
\end{figure}

We now study the population evolution for a QD with $\omega_0=4.16$ eV at $D=1$ nm, with 
$\tau_{eff}\approx$ 10 fs ($\tau_{eff}^{rad}\approx6$ fs and $\tau_{eff}^{tan}\approx14$ fs), taking into account that $\lambda_\perp(4.16,1)\approx 6.54\cdot10^5$ and 
$\lambda_\parallel(3.84,1)\approx 2.83\cdot10^5$, as shown in Fig.  \ref{fig2}. In Fig. \ref{fig7}, we present 
the time evolution of $|c_1(t)|^2$  (=$|c_2(t)|^2$ here) in case of a SIS (black solid curve) and an AIS (red dashed curve). We observe that the population evolution is different for the two initial states, which again points 
to quantum interference effects in the underlying dynamics. As in case of a QD with $\omega_0=3.84$ eV, shown 
in Fig. \ref{fig3}, the population dynamics shows a moderate non-Markovian character, and, at early times, is practically a decay with $\tau_{eff}$, while after $\approx$ 200 fs, the  population of the 
upper states is (almost) completely lost. We note that the population dynamics of $\ket{1}$ and $\ket{2}$ in 
case of the initial state $\ket{\Psi(0)}=\ket{1}$ (not shown here) looks analogous to the corresponding dynamics 
in case of a QD with  $\omega_0=3.84$ eV at $D=1$ nm, presented in the top panel of Fig. \ref{fig3} above.

\subsection{J-AGR dynamics at $\omega_0=4.16$ eV}

We now study the population dynamics of a J-AGR, i.e. a QE with $\tau_0=70$ ps, with $\omega_0=4.16$ eV
at $D=1.55$ nm, with $\tau_{eff}\approx$ 0.35 fs ($\tau_{eff}^{rad}\approx0.2$ fs and 
$\tau_{eff}^{tan}\approx0.5$ fs), taking into account that $\lambda_\perp(4.16,1)\approx 3.21\cdot10^5$ 
and $\lambda_\parallel(4.16,1.55)\approx 1.39\cdot10^5$, as shown in Fig.  \ref{fig2}. Here, the averaged 
effective decay rate, as well as the radial and tangential effective rates, are in attosecond time range, 
which is much smaller than the duration of the dynamics observed; we thus conclude that they are practically of minor importance for understanding the underlying strongly non-Markovian dynamics in this case.

\vspace*{1.cm}
\begin{figure}[h]
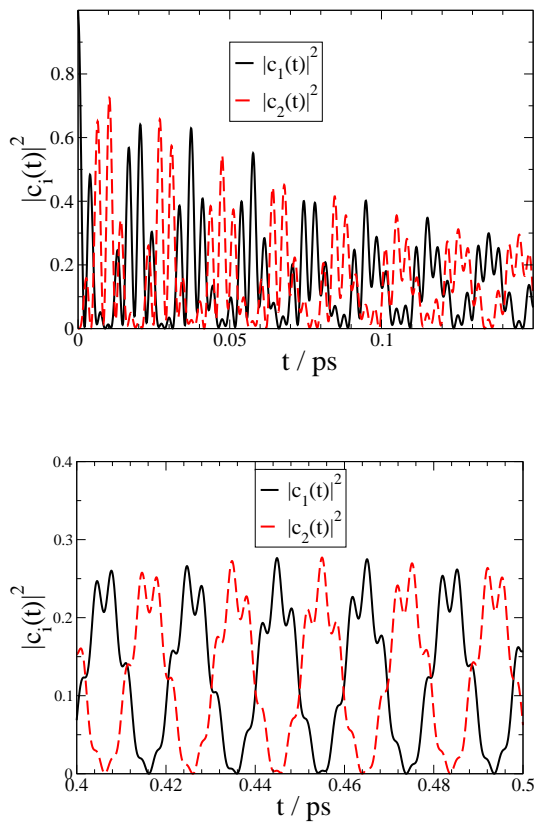

\centerline{\hbox{\epsfxsize=70mm\epsffile{fig8a.eps}}}
\vspace*{1.cm}
\centerline{\hbox{\epsfxsize=70mm\epsffile{fig8b.eps}}}
\begin{center}
\caption{\label{fig8} (color online) Top: Population evolution of $\ket{1}$ (black solid curve) and $\ket{2}$ (red  dashed curve) of a J-AGR with $\omega_0=4.16$ eV at $D=1.55$ nm up to 100 fs. Bottom: Population evolution of 
$\ket{1}$ (black solid curve) and $\ket{2}$ (red  dashed curve) of a J-AGR with $\omega_0=4.16$ eV at $D=1.55$ nm, at later times (400-500 fs).}
\end{center}
\end{figure}

In the top panel of Fig. \ref{fig8}, the time evolution of $|c_1(t)|^2$ (black solid curve) and
$|c_2(t)|^2$ (red dashed curve) are shown for the initial state $\ket{\Psi(0)}=\ket{1}$ during the first 150 fs. 
The strongly non-Markovian character of the underlying dynamics manifests itself as an oscillatory exchange of population between states $\ket{1}$ and $\ket{2}$ with the modified EM mode continuum as an intermediate, as clearly indicated by the phase difference of $\pi$ between $|c_1(t)|^2$ and $|c_2(t)^2$. Furthermore, the 
initial population of the {\sf V}-type system is decreased to about 40\% of its initial value within the first 150 fs.

\vspace*{1.cm}
\begin{figure}[h]
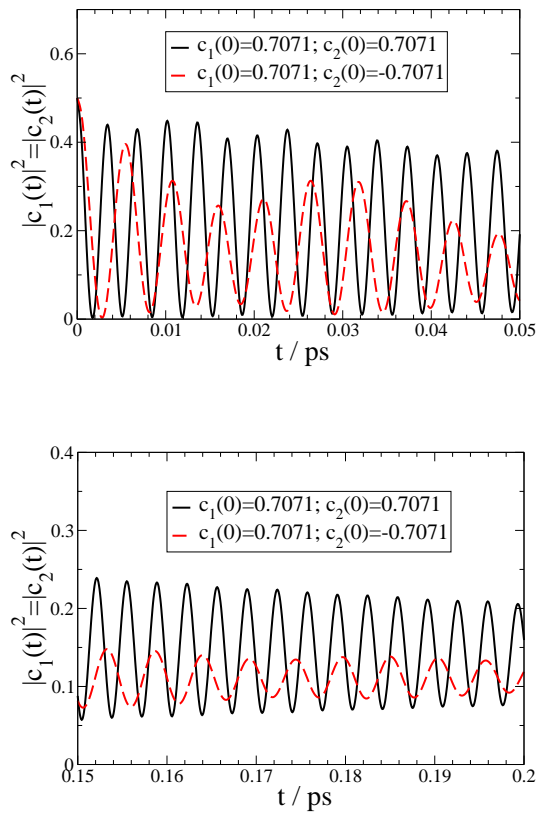

\centerline{\hbox{\epsfxsize=70mm\epsffile{fig9a.eps}}}
\vspace*{1.cm}
\centerline{\hbox{\epsfxsize=70mm\epsffile{fig9b.eps}}}
\begin{center}
\caption{\label{fig9} (color online) Top: Population evolution of $\ket{1}$  of the
J-AGR  with $\omega_0=4.16$ eV at $D=1.55$ nm for the initial states, $\ket{\Psi(0)}=\sqrt{0.5}\ket{1}+\sqrt{0.5}\ket{2}$ (black  solid curve) and $\ket{\Psi(0)}=\sqrt{0.5}\ket{1}-\sqrt{0.5}\ket{2}$
(red dashed curve) up to 50 fs. Bottom: Population evolution of $\ket{1}$  of the
J-AGR  with $\omega_0=4.16$ eV at $D=1.55$ nm for the initial states, $\ket{\Psi(0)}=\sqrt{0.5}\ket{1}+\sqrt{0.5}\ket{2}$ (black  solid curve) and $\ket{\Psi(0)}=\sqrt{0.5}\ket{1}-\sqrt{0.5}\ket{2}$
(red dashed curve) at later times (150-200 fs).}
\end{center}
\end{figure}

In the bottom panel of Fig. \ref{fig8} we show $|c_1(t)|^2$ (black solid curve) and $|c_2(t)|^2$ (red dashed curve) at later times, 400 - 500 fs. We now observe that the population evolution has reached a steady state;  
now, the non-decaying exchange of population between the {\sf V}-type system and the modified 
EM mode continuum amounts in total to $\approx$ 25\% of the initial population. Interestingly, contrary 
to the findings when studying the analogous configuration at $\omega_0=3.84$ eV shown in Fig. \ref{fig5}, 
the $\pi$-phase difference between $|c_1(t)|^2$ and $|c_2(t)|^2$, which existed at earlier times of the dynamics, is still present; here, clearly the population exchange between the two states of the {\sf V}-type 
system  with the modified EM continuum as an intermediate, suffers no dephasing, even over long time periods, 
in this case.

In Fig. \ref{fig9}, we investigate the influence of the initial state on the population evolution of states $\ket{1}$ and $\ket{2}$ for a J-AGR with $\omega_0=4.16$ eV at $D=1.55$ nm. In the top panel of this figure, we present
the time evolution of $|c_1(t)|^2$ and $|c_2(t)|^2$ for a SIS (black solid curve) and an AIS (red dashed curve) for the first 50 fs. We find that the population evolution of states $\ket{1}$ and $\ket{2}$ are strongly dependent on the initial state, which indicates strong quantum interference effects in the underlying non-Markovian dynamics.   In case of a SIS, the $|c_1(t)|^2$ ($=|c_2(t)|^2$ here) evolution shows a slowly decreasing 
in magnitude oscillatory transfer of population between the {\sf V}-type system and the modified EM mode continuum. In case of an AIS, the situation is similar, although the rate of magnitude decrease is faster. At later times, as shown in the bottom panel of Fig. \ref{fig9},  a moderate oscillatory population exchange, of about 20\% and 10\% of the initial population, in case of a SIS and an AIS, respectively, between states $\ket{1}$ and $\ket{2}$ and the modified EM modes occurs simultaneously. These minor qualitative differences on the population evolution of $\ket{1}$ and $\ket{2}$ are safely attributed to quantum interference effects in the underlying 
non-Markovian dynamics. Most interestingly, however, for both initial states, at later times, the population 
evolution shows clearly that about 20\% of the initial population is ``trapped'' in the {\sf V}-type system at all times, which indicates that coherent population trapping \cite{cpt} under conditions of strong coupling between 
a QE and a MNP, as studied here, is observable.

\subsection{The influence of the FCA on the dynamics of the {\sf V}-type QE: 
$K^\pm(\tau)$ vs $K^\pm_{FCA}(\tau)$}

\begin{figure}[h]
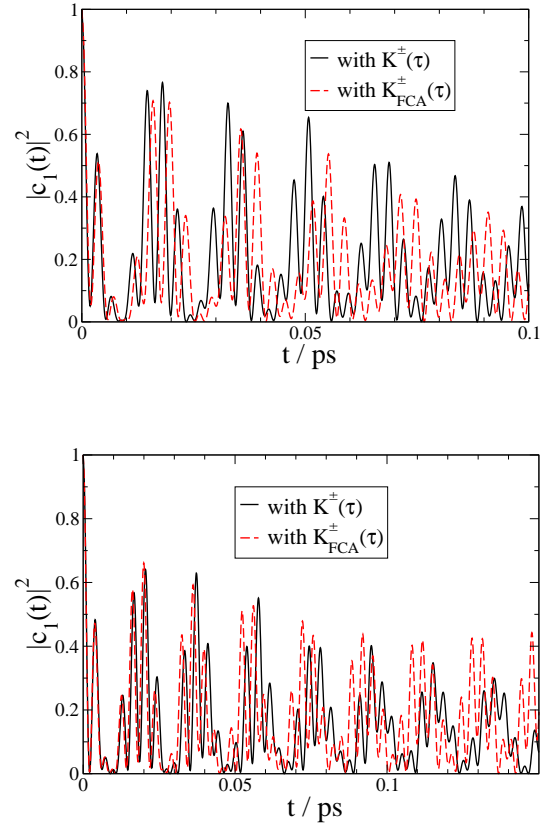

\centerline{\hbox{\epsfxsize=70mm\epsffile{fig10a.eps}}}
\vspace*{1.cm}
\centerline{\hbox{\epsfxsize=70mm\epsffile{fig10b.eps}}}
\begin{center}
\caption{\label{fig10} (color online) Population evolution of $\ket{1}$  of a QE  with $\tau_0=70$ ps
at $D=1.55$ nm obtained using the kernel $K^\pm(\tau)$  (defined by Eq. (\ref{with})) (black solid curve) and
using the kernel $K^\pm_{FCA}(\tau)$ (defined by Eq. (\ref{without})) (red dashed curve), with 
$\omega_0=3.84$ eV (top) and $\omega_0=4.16$ eV (bottom). See text for discussion.}
\end{center}
\end{figure}

We now investigate the influence of the FCA on the non-Markovian spontaneous emission dynamics of a 
{\sf V}-type QE. In the top panel of Fig. \ref{fig10}, we present the population evolution of state $\ket{1}$ 
with initial state $\ket{\Psi(0)}=\ket{1}$, for a J-AGR  at $D=1.55$ nm with $\omega_0=3.84$ eV, obtained 
using the $K^\pm(\tau)$ kernel (black solid curve), defined by Eq. (\ref{with}), and using the 
$K^\pm_{FCA}(\tau)$ kernel (red dashed curve), defined by Eq. (\ref{without}). We observe that the FCA introduces a phase shift between the two $|c_1(t)|^2$ curves. More specifically, the $|c_1(t)|^2$ obtained 
with $K^\pm_{FCA}(\tau)$ precedes the corresponding population evolution when the FCA is not invoked; we 
denote such a phase shift between the two curves as positive. In the bottom panel of Fig. \ref{fig10}, we present the population dynamics of state $\ket{1}$ for a J-AGR, again at $D=1.55$ nm, but now with $\omega_0=4.16$ eV, obtained using the $K^\pm(\tau)$ kernel (black solid curve) and with the $K^\pm_{FCA}(\tau)$ kernel (red dashed curve). We once more find that the two $|c_1(t)|^2$ curves are phase-shifted to each other. However,  interestingly, in this case, the phase shift is negative.

\vspace*{1.cm}

In order to better understand the phase shift in the population dynamics introduced by invoking the FCA, as discussed above, we further focus on the radial spectral density $J^{rad}(\omega)$ and
$J^{rad}_{FCA}(\omega)$, with and without the FCA, respectively, to be used in the corresponding kernels.
We do not elaborate on the case of the tangential spectral density here, since one comes to analogous 
findings as for the radial spectral case, to be presented now.

\begin{figure}[h]
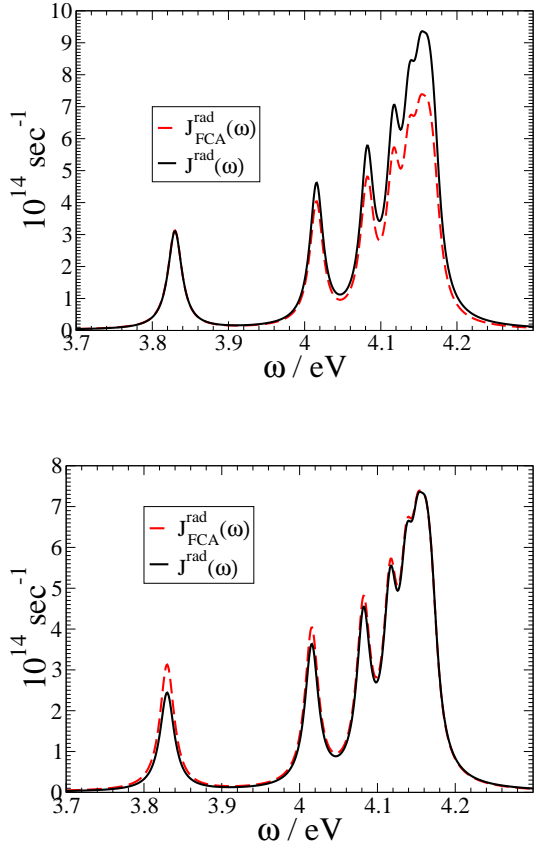

\centerline{\hbox{\epsfxsize=70mm\epsffile{fig11a.eps}}}
\vspace*{1.cm}
\centerline{\hbox{\epsfxsize=70mm\epsffile{fig11b.eps}}}
\begin{center}
\caption{\label{fig11} (color online) The radial spectral density $J^{rad}(\omega)$ (black solid curve)
and  $J^{rad}_{FCA}(\omega)$ (red dashed curve) for a QE  with $\tau_0=70$ ps at $D=1.55$ nm with 
$\omega_0=3.84$ eV (top) and $\omega_0=4.16$ eV (bottom). See text for discussion.}
\end{center}
\end{figure}

In Fig. \ref{fig11}, we show the  $J^{rad}(\omega)$ (black solid curve) and  $J^{rad}_{FCA}(\omega)$
(red dashed curve) in case of a QE with $\omega_0=3.84$ eV (top panel) and $\omega_0=4.16$ eV (bottom panel).
We observe that in both cases, around the resonance frequency $\omega_0$, $J^{rad}(\omega)$ and
$J^{rad}_{FCA}(\omega)$ are practically the same, which implies that in the common case of a (radial)
spectral density dominated by only one plasmonic resonance, or more generally, by several non-overlapping resonances, calculating the non-Markovian dynamics of a QE in the proximity of a MNP, with or without the FCA, leads to practically indistinguishable results. However, when the (radial) spectral density is dominated by 
overlapping plasmonic resonances, as in this work (see Fig. \ref{fig2}), invoking the FCA leads to observable differences between the two spectral densities as one moves away from the resonance frequency of the QE. 
In particular, in presence of overlapping resonances in the spectral density, $J^{rad}_{FCA}(\omega) < J^{rad}(\omega)$ is valid when $\omega > \omega_0$, with reversed relation between the two spectral densities, 
when  $\omega <\omega_0$. Furthermore, with help of Fig. \ref{fig11}, we can rationalise qualitatively the phase difference in the evolution of $|c_1(t)|^2$, when obtained with $K^\pm_{FCA}(\tau)$, instead of $K^\pm(\tau)$. We conclude that the introduction of the FCA for computing the population dynamics of a QE near a MNP, for 
which the spectral density is dominated by overlapping plasmonic resonances, introduces a positive (negative) 
phase shift in the population evolution, when the QE resonance frequency lies in the lower (higher) end of its 
non-vanishing spectral density.

\vspace*{1.cm}

\section{Dynamics of a two-level QE near a MNP}
\label{twolev}

In this section, we consider a {\sf V}-type QE, for which, however, the transition dipole moments $\vec{\mu}_{10}$
and $\vec{\mu}_{20}$ are along the $z$ and the $x$ axis, i.e. we use $\vec{\mu}_{10}=\mu\hat{\epsilon}_{z(x)}$
and $\vec{\mu}_{20}=\mu\hat{\epsilon}_{x(z)}$ in Eq. ({\ref{kernel}). In such a case, since $K^{ij}(\tau)=0$ 
($i\ne j$), the population dynamics of one excited state of the {\sf V}-type system does not depend on the 
dynamics of the other excited state; therefore one has effectively two independent two-level systems, instead of 
a three-level {\sf V}-type system. We also note that the two-level transition dipole polarisation  $\vec{\mu}$ along the $z$ (along the $x$ axis) implies that the population dynamics of such a two-level system is affected by the radial (tangential) spectral density only.

\vspace*{1.cm}
\begin{figure}[h]
\centerline{\hbox{\epsfxsize=70mm\epsffile{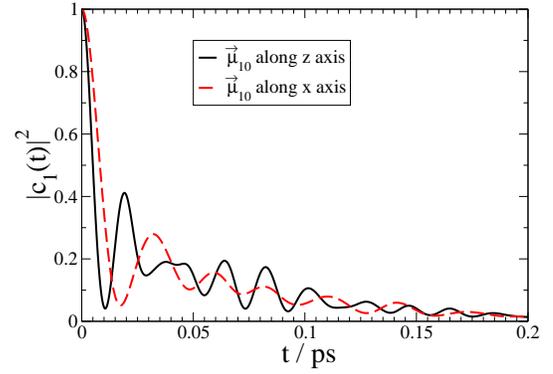}}}
\begin{center}
\caption{\label{fig12} (color online) Population evolution of $\ket{1}$ of a two-level system with
$\omega_0=4.16$ eV and $\tau_0=4$ ns at $D=1$ nm with $\vec{\mu}_{10}=\mu\hat{\epsilon}_z$
(black solid curve) and  $\vec{\mu}_{10}=\mu\hat{\epsilon}_x$  (red dashed curve). See text for discussion.}
\end{center}
\end{figure}

We now study the dynamics for such a two-level system in order to compare its population dynamics to the 
dynamics of a {\sf V}-type system presented above. For the sake of concreteness, we focus on a two-level 
QE, composed of states $\ket{0}$ and  $\ket{1}$ (see Fig. \ref{fig1}), with resonance frequency 
$\omega_0=4.16$ eV; the results at a lower $\omega_0$ (not shown here) are similar to the results at 
$\omega_0=4.16$ eV, with only minor quantitative differences due to the much smaller enhancement factors 
$\lambda^\perp(\omega_0,D)$  and $\lambda^\parallel(\omega_0,D)$, when $\omega_0 < 4.16$ eV.

In Fig. \ref{fig12}, we present the population dynamics of state $\ket{1}$ of a two-state QE with $\tau_0=4$ ns
at $D=1$ nm, with a transition dipole moment along the $z$ axis (black solid curve) and along the $x$ axis (red dashed curve); the corresponding effective decay times $\tau_{eff}^{rad}$ and $\tau_{eff}^{tan}$ are 
6 fs and 14 fs, respectively, taking into account that $\lambda^\perp(4.16,1)\approx 6.54\cdot10^5$ and $\lambda^\parallel(3.84,1)\approx 2.83\cdot10^5$ as shown in Fig.  \ref{fig2}. The time evolution of $|c_1(t)|^2$ 
at early times can be understood as a decay with the corresponding effective decay time. However, quickly, 
an oscillatory decay of the population of $\ket{1}$ into the modified EM mode continuum with distinctly 
non-Markovian character dominates the dynamics. Moreover, an interesting conjecture regarding the population dynamics shown in this figure can be made, when a comparison with the population dynamics for the {\sf V}-type system with $\omega_0=4.16$ eV and $\tau_0=4$ ns at $D=1$ nm presented in Fig. \ref{fig7} above is made. 
More specifically, we observe that the curves of $|c_1(t)|^2$ in Fig. \ref{fig7} and in Fig. \ref{fig12} look very similar, when taking into account the different amount of initial population, although they correspond to distinctly different initial conditions. The comparison of the results of these two figures suggests strongly that the 
different population evolution of $|c_1(t)|^2$ for the SIS and the AIS shown in Fig. \ref{fig7} is due to the 
fact that for a SIS state, due to quantum interference, the dynamics is dominated by the spectral density along 
the $z$ axis, i.e. the radial spectral density. In case of an AIS, the $|c_1(t)|^2$ evolution in Fig. \ref{fig7} is dominated by the tangential spectral density, i.e. along the $x$ axis \cite{paspa11}. We can thus understand  
better, with help of the two-level system studied here, the direct result of the quantum interference on the dynamics of the {\sf V}-type system with $\omega_0=4.16$ eV and $\tau_0=4$ ns at $D=1$ nm, discussed 
above.

\vspace*{1.cm}
\begin{figure}[h]
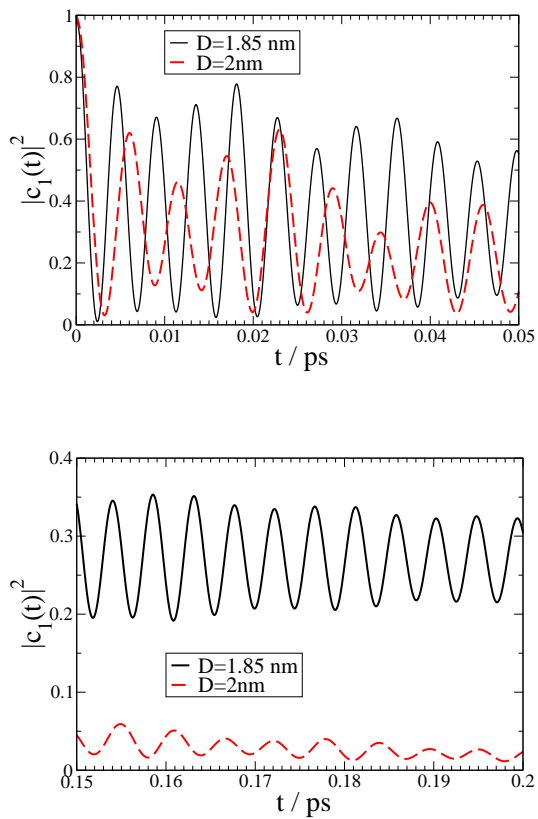

\centerline{\hbox{\epsfxsize=70mm\epsffile{fig13a.eps}}}
\vspace*{1.cm}
\centerline{\hbox{\epsfxsize=70mm\epsffile{fig13b.eps}}}
\begin{center}
\caption{\label{fig13} (color online) Population evolution of $\ket{1}$ of a two-level system with
$\tau_0=70$ ps and transition dipole moment along the $z$ axis at $D=1.85$ nm (black solid curve) 
and at $D=2$ nm (red dashed curve) up to 100 fs (top panel), and at later times, 200-300 fs (bottom panel).
See text for discussion.}
\end{center}
\end{figure}

In Fig. \ref{fig13}, we consider a two-level QE with $\omega_0=4.16$ eV and $\tau_0=70$ ps with a transition dipole moment along the $z$ axis, which implies that the dynamics is affected only by the radial spectral density. 
In this figure we present the population evolution $|c_1(t)|^2$ at $D=1.85$ nm (black solid curve) and
at $D=2$ nm (red dashed curve) up to 50 fs (top panel) and at later times, 150-200 fs (bottom panel). Here, 
the radial effective decay times $\tau_{eff}^{rad}$ and  $\tau_{eff}^{tan}$ are about 0.5 fs and 1.5 fs, respectively, taking into account that $\lambda^\perp(4.16,1.85)\approx 1.39\cdot10^5$ and 
$\lambda^\perp(4.16,2)\approx 4.86\cdot10^4$ as shown in Fig. \ref{fig2}.

In the top panel of Fig. \ref{fig13}, the population evolution of state $\ket{1}$ at $D=1.85$ nm and at $D=2$ nm
shows strong non-Markovian character; only at very early times, the decay dynamics is dictated by 
the corresponding $\tau_{eff}^{rad}$. After 50 fs, at $D=1.85$ nm, the population of state $\ket{1}$
is decreased to about half its initial value, which resembles the time evolution of $|c_1(t)|^2$ with an AIS
of a {\sf V}-type system with resonance frequency 4.16 eV and free-space decay time 70 ps at $D=1.55$ nm,
shown in Fig. \ref{fig9}. We thus conclude that the coupling conditions of the two-level QE here are comparable  
to the coupling conditions of the {\sf V}-type QE related to the results shown in Fig. \ref{fig9}; a comparative discussion of the population dynamics shown in these two figures is then justifiable. With help of the results of 
the two-level system shown in Fig. \ref{fig13}, we can now conclude that the evolution of $|c_1(t)|^2$ of a 
{\sf V}-type system with an AIS, as shown in the top panel of Fig. \ref{fig9}, is dominated primarily by the 
radial spectral density of the modified EM modes.

In the bottom panel of Fig. \ref{fig13}, we find that at $D=1.85$ nm, at later times, there is a steady population 
exchange of about 15\% of the initial population between state $\ket{1}$ and the modified radial EM modes, 
while at $D=2$ nm, the corresponding population exchange amounts only to about 5\%. Most 
importantly, at $D=1.85$ nm, the population exchange occurs while about 20\% of the initial population remains 
in the two-level QE at all times. At $D=2$ nm, however, the initial population of state $\ket{1}$ is almost 
completely transferred in the EM continuum, with clear indication that it will ultimately wane out totally. 
We thus conclude that population trapping is also observable for a two-level QE near a MNP under the given conditions.

We have also studied the influence of the FCA in the dynamics of a two-level system (not shown here); our conclusions are similar to the case of a {\sf V}-type system discussed above. 

\section{Conclusions}
\label{concl}

In conclusion, we have studied the non-Markovian quantum dynamics for various initial states of a degenerate
{\sf V}-type QE near a MNP, which exhibits anisotropic Purcell effect. We have considered QEs with free-space decay time in the ns and ps time regime, for various resonance frequencies, corresponding to moderate ($10^4$) 
and strong ($10^5$) enhancement of the free-space decay rate. We observe a transition in the upper states
population time evolution, from a gradual decay of the total population to a steady oscillatory population exchange between the QE and the modified EM modes. This effect occurs when strong enhancement of the free-space spontaneous decay rate of the QE occurs, and, in some cases it can lead to coherent population trapping in the
{\sf V}-type system. Furthermore, the strong dependence of the population dynamics on a particular initial state, 
at otherwise identical conditions, clearly indicates that spontaneous emission interference of the QE excited states into the modified EM modes continuum takes place.  We have also studied a two-level QE near a MNP at the same
conditions as the {\sf V}-type system, coming to similar conclusions as for the three-level system.

Lastly, we have investigated the influence of the FCA on the calculated population dynamics for the above three- and two-level systems. In both cases, we conclude that the FCA can affect the quality of the calculated population dynamics only in systems, for which the modified by the MNP spectral density of the EM modes is dominated by overlapping plasmonic resonances. In such cases, the FCA introduces a positive (negative) phase shift in the population time evolution with respect to the exact dynamics without invoking the FCA; a positive (negative) phase shift appears when the resonance frequency of the QE lies in the low (high) end of the spectral density. In case of spectral densities with only non-overlapping plasmonic resonances, however, the FCA is an excellent approximation.
We believe that our findings can be particularly useful in the development of quantum technological applications involving plasmonic nanostructures \cite{zhang15}.

\section*{Acknowledgments}

E.P. acknowledges the support of ``Research Projects for Excellence IKY/Siemens''.

\end{document}